\documentclass[12]{iopart}

\usepackage{iopams}
\usepackage{amsfonts,bbm}
\usepackage[mathscr]{eucal}
\usepackage{epsfig}

\newcommand{\Tra}{{\rm tr}\,}

\newcommand{\ket}[1]{\left|{#1}\right\rangle}

\newcommand{\ketbrad}[1]{\left|{#1}\rangle\!\langle{#1}\right|}

\newcommand{\mean}[1]{\langle{#1}\rangle}

\newcommand{\rank}[1]{{\rm rank}\left(#1\right)}
\renewcommand{\vec}[1]{\mbox{\boldmath$#1$}}

\begin{document}
\title{Decomposition of any quantum measurement into extremals}

\author{G.~Sent\'{i}s}
\address{F\'isica Te\`orica: Informaci\'o i Fen\`omens Qu\`antics, Universitat
Aut\`{o}noma de Barcelona, 08193 Bellaterra (Barcelona), Spain}
\author{ B.~Gendra}
\address{F\'isica Te\`orica: Informaci\'o i Fen\`omens Qu\`antics, Universitat
Aut\`{o}noma de Barcelona, 08193 Bellaterra (Barcelona), Spain}
\author{S. D. Bartlett}
\address{Centre for Engineered Quantum Systems, School of Physics, The University of Sydney, Sydney, NSW 2006, Australia}%
\author{A. C. Doherty}
\address{Centre for Engineered Quantum Systems, School of Physics, The University of Sydney, Sydney, NSW 2006, Australia}%

\begin{abstract}
We design an efficient and constructive algorithm to decompose any generalized quantum measurement into a convex combination of extremal measurements. 
We show that if one allows for a classical post-processing step only extremal rank-1 POVMs are needed. 
For a measurement with $N$ elements on a $d$-dimensional space, our algorithm will decompose it into at most $(N-1)d+1$ extremals, whereas the best previously known upper bound scaled as $d^2$.
Since the decomposition is not unique, we show how to tailor our algorithm to provide particular types of decompositions that exhibit some desired property.
\end{abstract}
\pacs{03.67.Hk, 03.65.Ta}

\maketitle

\section{Introduction}

The growth of quantum information theory and, in particular, the development of a vast variety of quantum processing techniques in the past few decades has drawn major attention towards the measurement process in quantum mechanics. Because no complete knowledge of the state of a quantum system can be retrieved from a single measurement, in general there are different incompatible measurement strategies that may yield very different results when applied to the same scenario. Hence, most often the design of a quantum processing technique involves finding which measurement best accomplishes a specific task, or which sequence of measurements is statistically optimal. These problems are the keystone of quantum estimation theory \cite{Helstrom1969}, and its solutions stand as a characteristic feature of many quantum processing tasks. 

Recent advances in experimental techniques 
%had helped many of these tasks to be realizable in a lab, 
have rendered many of these tasks realizable in a laboratory,
where a minimum resource perspective prevails. The sought for the minimum resources needed to implement a certain task has a paradigmatic example in quantum state preparation: to prepare all pure states of a bipartite system, it is enough to prepare only one maximally entangled pure state; then, by means of local operations and classical communication, one can obtain any bipartite pure state \cite{Nielsen2000}. The mathematical object that represents a general quantum measurement is a \textit{positive operator valued measure} (POVM), and therefore these kind of questions concern to the mathematical structure of POVMs. The aim of this paper is to address the following minimum resource problem: 
%given a certain POVM, what are the simplest resources needed to implement it, and how should one use them?
given a certain POVM, what are the simplest resources needed, and how one can implement it in terms of them?

POVMs form a convex set. This means that, given two known POVMs, any randomized implementation of them is also a POVM: just as mixed states are probabilistic mixtures of pure states, one can talk about measurements that can be regarded as probabilistic mixtures of POVMs. Those that cannot be expressed as combinations of other measurements are called extremal POVMs. Since many measurement optimization problems consist in maximizing a convex figure of merit, which leads to an extremal solution, this type of POVM appears quite frequently. It is no wonder then that the characterization of extremal POVMs has been extensively addressed in the literature %when these act on Hilbert spaces of both finite and infinite dimension
\cite{D'Ariano2005,Chiribella2010,Heinosaari2011,Pellonpaa2011}. 

It is clear that the set of all extremal POVMs comprise the toolbox needed to effectively implement any measurement, as an appropriate convex combination of extremal POVMs will reproduce its statistics. A number of works have been devoted to prove the existence of such decompositions of measurements into extremals for finite \cite{D'Ariano2005,Haapasalo2011} as well as infinite dimensional systems \cite{Chiribella2007b}. 
%In particular, D'Ariano \textit{et al.} give in \cite{D'Ariano2005} a generic algorithm for decomposing a point in a convex set into a combination of extremal points of that set that can be applied to POVMs. Nonetheless, we consider that the question of which are the minimal resources one needs to implement a given POVM yet remains unclear from an operational point of view.
However, the question of which are the minimal resources needed to implement a given POVM remains unclear from an operational point of view.
In this paper we provide a clear answer to this question by designing a constructive and efficient algorithm that takes as input any POVM with an arbitrary (but finite) number of outcomes and gives as output a convex combination of extremal POVMs that reproduces its statistics. We show that only rank-1 extremal POVMs are needed if one allows for a classical post-processing of the outcomes (in agreement to a similar result shown in \cite{Haapasalo2011}). 
The number of extremals that this algorithm produces is upper bounded by $(N-1)d+1$, where $N$ is the number of outcomes of the input POVM and $d$ is the dimension of its associated Hilbert space. This bound is significantly lower than the best previously known upper bound \cite{D'Ariano2005}, which scaled as $d^2$.
%Given a POVM with $N$ outcomes on a $d$-dimensional space, our algorithm yields a number of extremals which is upper bounded by $(N-1)d+1$.
As a byproduct of our analysis, we obtain a simple geometrical characterization of extremal POVMs in terms of the generalized Bloch vectors associated to their elements.

In Section \ref{notation} we fix the notation and illustrate how the algorithm works in a few simple cases. In Section \ref{geometric} we set the mathematical tools we rely on and we derive from them a geometrical characterization of extremal POVMs. Section \ref{algorithm} is devoted to the full description of the algorithm, and Section \ref{ordereddecomp} to the discussion of further improvements. We finally summarize our results.

\section{Simple cases}\label{notation}

%Let us start by fixing the notation and conventions used throughout this paper. We will focus on POVMs with a finite number of outcomes on a Hilbert space $\mathcal{H}$ of dimension $d$. A POVM with $N$ outcomes is a set $\mathbb{P}_N=\{E_i\}_{i=1}^N$ of positive semidefinite operators on $\mathcal{H}$, which satisfy the normalization condition $\sum_i E_i =\mathbb{I}$. The operators $E_i$ are called {\it POVM elements}. 
%%
%A convex combination of two POVMs is also a POVM: suppose that $\mathbb{P}_3^{(1)}=\left\{E_1,E_2,E_3\right\}$ and $\mathbb{P}_3^{(2)}=\left\{E_3,E_4,E_5\right\}$ are two 3-outcome POVMs, then
%$\mathbb{P}_5 \equiv p_1\mathbb{P}_3^{(1)} + p_2\mathbb{P}_3^{(2)} = \left\{p_1E_1,p_1E_2,(p_1+p_2)E_3,p_2E_4,p_2E_5\right\}$ is also a POVM, where $p_1+p_2=1$. The convex combination $\mathbb{P}_5$ contains the weighted union of $\mathbb{P}_3^{(1)}$ and $\mathbb{P}_3^{(2)}$, but identical outcomes are binned together. 
%%

Let us start by fixing the notation and conventions used throughout this paper. A POVM is a set $\mathbb{P}=\{E_i\}$ of positive semidefinite operators acting on a Hilbert space $\mathcal{H}$ of dimension $d$, which satisfy the normalization condition $\sum_i E_i =\mathbb{I}$. The operator $E_i$ is called a {\it POVM element}, and it is associated to the outcome $i$ of the POVM.
In this paper we focus on POVMs with a finite number of outcomes.
The elements $E_i$ might be zero for some $i$, meaning that the corresponding outcomes have zero probability of occurrence. Two POVMs that differ only in the number or position of their zero elements are considered to be physically equivalent. When characterizing a POVM by its number of outcomes we will refer only to those with physical meaning, that is to the outcomes with a  non-zero operator associated. In this spirit, we denote by $\mathbb{P}_N$ a POVM $\mathbb{P}$ with $N$ non-zero elements, and we will refer to it as a $N$-outcome POVM.

A convex combination of two POVMs is also a POVM: suppose that $\mathbb{P}_3^{(1)}=\left\{E_1,E_2,E_3,0,0\right\}$ and $\mathbb{P}_3^{(2)}=\left\{0,0,E_3,E_4,E_5\right\}$ are two 3-outcome POVMs, then
$\mathbb{P}_5 \equiv p_1\mathbb{P}_3^{(1)} + p_2\mathbb{P}_3^{(2)} = \left\{p_1E_1,p_1E_2,(p_1+p_2)E_3,p_2E_4,p_2E_5\right\}$ is also a POVM, where $p_1+p_2=1$. The convex combination $\mathbb{P}_5$ is the weighted sum element-by-element of $\mathbb{P}_3^{(1)}$ and $\mathbb{P}_3^{(2)}$.

\begin{figure}[t]
\includegraphics[scale=1.5]{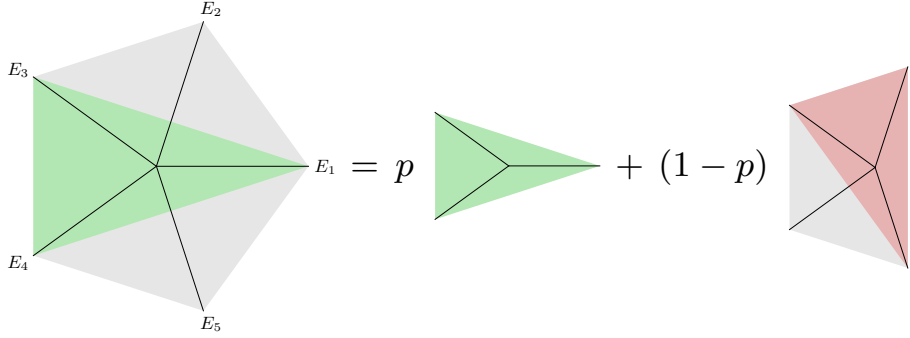}
\caption{(Color online) First step of the decomposition of $\mathbb{P}_5$. The selection of elements (green) form the trine $\mathbb{P}_3^{(1)}$ which appears in the decomposition with associated probability $p$. After extracting it, we are left with $\mathbb{P}_4^{({\rm aux})}$ with associated probability $(1-p)$. In the second step we select another trine (red) from $\mathbb{P}_4^{({\rm aux})}$.}\label{fig1}
\end{figure}

In this paper we are faced with the reverse situation: given a POVM, we want to find a decomposition into a convex combination of smaller (i.e. with less outcomes) POVMs. As a simple example of this type of decomposition, consider the POVM needed in the eavesdropping of the ``BB84'' protocol \cite{Nielsen2000} 
\begin{equation}
  \mathbb{P}_4 =
  \left\{{\footnotesize{1\over2}} \ketbrad{0},{\footnotesize{1\over2}} \ketbrad{1},
  {\footnotesize{1\over2}} \ketbrad{+}, {\footnotesize{1\over2}}\ketbrad{-}
  \right\}\,.
\end{equation}
Note that $\mathbb{P}_4$ can be expressed as
\begin{equation}
  \mathbb{P}_4 = {\footnotesize{1\over2}}\mathbb{P}_2^{(z)} + {\footnotesize{1\over2}}\mathbb{P}_2^{(x)} \,,
\end{equation}
where
\begin{eqnarray}
  \mathbb{P}_2^{(z)} &= \left\{\ketbrad{0},\ketbrad{1},0,0\right\} \\
  \mathbb{P}_2^{(x)} &= \left\{0,0,\ketbrad{+},\ketbrad{-}\right\}\,.
\end{eqnarray}
Thus, the POVM $\mathbb{P}_4$ can be effectively implemented by tossing an
unbiased coin, and then performing either $\mathbb{P}_2^{(x)}$ or
$\mathbb{P}_2^{(z)}$ based on the outcome of this toss.
In this case it is trivial to identify at sight the two pairs of orthogonal operators and their weights in the decomposition. This will not be so for an arbitrary measurement. The next example is presented to gain insight on how this operation can be performed algorithmically. Consider the POVM with five outcomes
\begin{equation}
\mathbb{P}_5=\left\{{\footnotesize{2\over 5}}E_1,{\footnotesize{2\over 5}}E_2,{\footnotesize{2\over 5}}E_3,{\footnotesize{2\over 5}}E_4,{\footnotesize{2\over 5}}E_5\right\}\,,
\end{equation}
where $E_i$ are rank-1 projectors lying on the equator of the Bloch sphere and aligned on the directions shown in Fig~\ref{fig1}. To carry out its decomposition, one first notices that some subsets of $\{E_i\}$ may form a smaller POVM by themselves with appropriate weights. Then, by selecting one of these subsets (for instance the trine formed by elements 1, 3 and 4), one can rewrite the original POVM as
\begin{equation}
\mathbb{P}_5=p \mathbb{P}_3^{(1)} + (1-p) \mathbb{P}_4^{({\rm aux})}\,,
\end{equation}
where $p=1/\sqrt{5}$ and
\begin{eqnarray}
&\mathbb{P}_3^{(1)} \;\;\;= \left\{ {\footnotesize{2\over \sqrt{5}}} E_1,0, \left(1-{\footnotesize{1\over \sqrt{5}}}\right) E_3, \left(1-{\footnotesize{1\over \sqrt{5}}}\right) E_4,0\right\} \,, \\
&\mathbb{P}_4^{({\rm aux})} = \left\{ 0,{\footnotesize{2\over 5-\sqrt{5}}} E_2, {\footnotesize{3-\sqrt{5}\over 5-\sqrt{5}}} E_3, {\footnotesize{3-\sqrt{5}\over 5-\sqrt{5}}} E_4, {\footnotesize{2\over 5-\sqrt{5}}} E_5\right\}\,.
\end{eqnarray}
Note that both $\mathbb{P}_3^{(1)}$ and $\mathbb{P}_4^{({\rm aux})}$ are strictly smaller POVMs than $\mathbb{P}_5$. The operation just performed consists in algebraically extracting $\mathbb{P}_3^{(1)}$, in such a way that the remaining elements form a POVM with at least one less outcome (in the following section we prove that such an operation is always possible). Note also that $\mathbb{P}_4^{({\rm aux})}$ is further decomposable. Proceeding iteratively, one can select this time the elements 2, 3 and 5 and write the convex combination of trines 
\begin{equation}
\mathbb{P}_4^{({\rm aux})}=p' \mathbb{P}_3^{(2)}+(1-p')\mathbb{P}_3^{(3)}\,,
\end{equation}
where $p'=1/2$ and
\begin{eqnarray}
&\mathbb{P}_3^{(2)} = \left\{ 0,\left(1-{\footnotesize{1\over \sqrt{5}}}\right) E_2, \left(1-{\footnotesize{1\over \sqrt{5}}}\right) E_3, 0,{\footnotesize{2\over \sqrt{5}}} E_5\right\} \,, \\
&\mathbb{P}_3^{(3)} = \left\{ 0,{\footnotesize{2\over \sqrt{5}}} E_2, 0, \left(1-{\footnotesize{1\over \sqrt{5}}}\right) E_4,\left(1-{\footnotesize{1\over \sqrt{5}}}\right) E_5\right\}\,.
\end{eqnarray}
Finally, the original 5-outcome POVM can be expressed as a convex combination of 3-outcome POVMs as
\begin{equation}
\mathbb{P}_5=p_1\mathbb{P}_3^{(1)}+p_2\mathbb{P}_3^{(2)}+p_3\mathbb{P}_3^{(3)}\,
\end{equation}
where $p_1=p$, $p_2=(1-p)p'$ and $p_3=(1-p)(1-p')$.

Note that both $\mathbb{P}_5$ and $\mathbb{P}_4$ in the previous examples are rank-1 POVMs\footnote{A POVM is called rank-1 iff all its non-zero elements $E_i$ are rank-1 operators, i.e. they can be written as $E_i=e_i P_i$, where $0< e_i\leqslant 1$ and $P_i$ is a normalized one-dimensional projector.}, and hence we need no more than convex combinations of rank-1 POVMs to implement them. However, consider the full-rank 2-outcome POVM
\begin{equation}
\mathbb{P}_2=\left\{{\footnotesize{1\over 2}} \ketbrad{0}, {\footnotesize{1\over 2}} \ketbrad{0}+\ketbrad{1}\right\} \, .
\end{equation}
In this case it will be enough to measure $\mathbb{P}_2^{(z)}=\{\ketbrad{0},\ketbrad{1}\}$ and, if its first outcome is obtained, then toss an unbiased coin to decide between the two outcomes of $\mathbb{P}_2$. The projector $\ketbrad{0}$, an element of $\mathbb{P}_2^{(z)}$, is associated with more than one element of $\mathbb{P}_2$, thus the association of the obtained outcome with an original outcome is not immediate. This situation requires an additional step: classical post-processing of the outcomes. This kind of operation has been previously introduced in the literature under the name of \textit{relabeling} \cite{Haapasalo2011}. In general, the post-processing step will be necessary whenever $\rank{\mathbb{P}_N}>1$. For any original element $E_i$ such that $\rank{E_i}>1$, we will split it into a combination of rank-1 operators (by writing it in its eigenbasis) and consider such operators as additional outcomes, thus obtaining a rank-1 POVM that is statistically equivalent to the original one. Of course, to reproduce the statistics accordingly, a map from such new outcomes to the original ones is needed.
We address in full detail the case of POVMs of higher rank and the inclusion of a post-processing step in Section \ref{algorithm}.

%We have seen in this Section examples of measurements that are mixtures of other measurements. The mathematical structure of POVMs is convex: any inner point of the set of POVMs corresponds to a mixed measurement, i.e. it can be expressed as a convex combination of two different POVMs. We denote by $\mathcal{P}_N$ the convex set of POVMs with $N$ outcomes on $\mathcal{H}$. Note that this convex set also contains any POVM with a strictly smaller number of outcomes: such POVM can still be considered as having $N$ outcomes, some of which have zero probability of occurrence. Hence, if $M\leqslant N$, then $\mathcal{P}_M \subset \mathcal{P}_N$. Obviously, $\mathcal{P}_N \subset \mathcal{P}$ for any value of $N$, where $\mathcal{P}$ is the convex set of all POVMs. 

We have seen in this Section examples of measurements that are mixtures of other measurements. The mathematical structure of POVMs is convex: any inner point of the set of POVMs corresponds to a mixed measurement, i.e. it can be expressed as a convex combination of two different POVMs. We denote by $\mathcal{P}_N$ the convex set of POVMs with $N$ elements on $\mathcal{H}$. Note that for any $\mathbb{P} \in \mathcal{P}_N$ we can construct a physically equivalent POVM $\tilde{\mathbb{P}} \in \mathcal{P}_M$, with $M \geqslant N$, just by adding zero-elements to $\mathbb{P}$. The limit of infinite elements yields the convex set of all POVMs $\mathcal{P}$.

An \textit{extremal} POVM is a measurement that cannot be expressed as a mixture of two other POVMs. The 2- and 3-outcome POVMs obtained in the examples above are extremal. If a POVM with $N$ elements $\mathbb{P}$ is extremal in the convex set $\mathcal{P}_N$, then any physically equivalent POVM with $M$ elements $\tilde{\mathbb{P}}$, $M\geqslant N$, is also extremal in $\mathcal{P}_M$. Ultimately, $\mathbb{P}$ will be associated with a set of extremal points of $\mathcal{P}$. %The set of all extremal POVMs corresponds to the boundary of the convex set $\mathcal{P}$. 
So far we have used an apparently more restricted definition of extremality. From the logic of the decompositions presented, it follows that we are considering a rank-1 POVM $\mathbb{P}_N=\{E_i\}$ to be extremal iff there does not exist any subset $\{E_k\}\subset\mathbb{P}_N$, $k=1,\ldots,M<N$ such that $\mathbb{P}_M=\{a_k E_k\}$ is itself a POVM for a suitable set of positive coefficients $\{a_k\}$. We have seen that if such a subset exists, then $\mathbb{P}_N$ can be split in $\mathbb{P}_M$ plus another POVM. We are therefore considering only decompositions into extremals formed by a subset of elements of the original $\mathbb{P}_N$. However, we prove in Section \ref{geometric} that looking for such subsets is sufficient to check for extremality of a given POVM.

\section{Selection of extremal POVMs and geometric characterization}\label{geometric}

The decomposition of the POVMs presented as examples above is achieved through the selection of subsets of their elements capable of forming a POVM by themselves. In order to give some insight on how to perform this selection for a general POVM $\mathbb{P}$ with $N$ outcomes, we now examine the conditions under which a set of $n$ arbitrary rank-1 operators $\{E_i\}$ can comprise a POVM, that is, there is a set of positive coefficients $\{a_i\}$ such that $\sum_{i=1}^{n} a_i E_i={\mathbb I}$. For simplicity and w.l.o.g. we will assume the operators $E_i$ to be normalized (i.e. $\Tra E_i=1$). Recall that, for a $d$-dimensional Hilbert space, we can express $E_i$ in a generalized Bloch-like representation as
\begin{equation}\label{elements}
E_i=\left(\frac{1}{d} {\mathbb I} + \frac{1}{2} \sum_j \mean{\hat{\lambda}_j}_i \hat{\lambda}_j \right) \, ,
\end{equation}
where the operators $\hat{\lambda}_j$, $j=1,\dots,d^2-1$ are an orthogonal basis of generators of SU$(d)$ and the generalized Bloch vector $\vec{v}_i$ is defined with their expectation values: \mbox{$\vec{v}_i \equiv (\mean{\hat{\lambda}_1}_i,\dots,\mean{\hat{\lambda}_{d^2-1}}_i)$}. In this representation, pure states have associated a generalized Bloch vector of fixed length $|\vec{v}|=\sqrt{2(d-1)/d}$. Then, the POVM condition may be equivalently written as

\begin{eqnarray}
\sum_i a_i = d \label{cond3} \, ,\\
\sum_i a_i \vec{v}_i = \vec{0} \label{cond4} \, ,
\end{eqnarray}
that is a system of $d^2$ linear equations. At this point we are only interested in checking the consistency of (\ref{cond3}) and (\ref{cond4}). Therefore, the existence of the set $\{a_i\}$ can be cast as a \textit{linear programming feasibility problem}. 

Before proceeding further, let us briefly overview the standard linear programming formalism (for an extensive review on the topic see e.g. \cite{Optimization2004,Todd2002}). A general \textit{linear program} (LP) has the standard form
\begin{eqnarray}\label{LP}
\min &\quad& c^T x \nonumber \\
\mbox{subject to} &\quad& Ax=b \nonumber \\
&\quad& x\geq 0 \, ,
\end{eqnarray}
where $A\in \mathbb{R}^{p \times q}$, $b\in \mathbb{R}^p$ and $c\in \mathbb{R}^q$ are the given data, and the vector $x\in \mathbb{R}^q$ is the variable to optimize. We call (\ref{LP}) \textit{feasible} if there exists $x\in \mathbb{R}^q$ such that $Ax=b$, $x\geq0$. Any LP of the standard form above has a \textit{dual problem} of the form
%Its dual problem has the form
%
\begin{eqnarray}\label{dual}
\max &\quad& -b^T \nu \nonumber \\
\mbox{subject to} &\quad& A^T \nu +c \geq 0 \, ,
\end{eqnarray}
where $\nu \in \mathbb{R}^p$. Let us assume that both LPs~(\ref{LP}) and (\ref{dual}) are feasible. Then, we may write
\begin{equation}\label{dualcheck}
c^T x + b^T \nu = x^T c + x^T A^T \nu = x^T (c+A^T \nu) \geq 0 \, .
\end{equation}
In order to obtain feasibility conditions of the LP (\ref{LP}), we now set $c=0$ and solve it. The existence of a solution implies that (\ref{LP}) is feasible and, from (\ref{dual}) and (\ref{dualcheck}), 
%that there exists a vector $\nu$ such that $A^T \nu \geq 0$ and $b^T \nu \geq 0$. 
that for all vectors $\nu$, $A^T \nu \geq 0$ implies $b^T \nu \geq 0$.
If the dual problem does not have a solution, then its corresponding LP neither has one. Conversely, the existence of a vector $\nu$ that verifies the conditions
\begin{eqnarray}
A^T \nu &\leq& 0 \label{dualcond1} \, , \\
b^T \nu &>& 0 \label{dualcond2} \, ,
\end{eqnarray}
implies the infeasibility of (\ref{LP}). Notice that finding a $\nu$ subject to $A^T \nu \geq 0$, $b^T \nu <0$ is an equivalent problem.

We are now in the position to reinterpret the problem of finding the set of coefficients $\{a_i\}$ within the general linear program scheme presented above. The components of the vector $x$ are the coefficients we want to determine, that is
$
x=\{a_1,a_2,\dots,a_n\} .
$
Conditions (\ref{cond3}) and (\ref{cond4}) can be cast together in the \mbox{$Ax=b$} equation: $A$ is a matrix whose columns are given by vectors $v_i=(\vec{v}_i,1)$, and $b=(\vec{0},d)$. Therefore, the dimensions of this linear program are given by $p\equiv d^2, q\equiv n$. In the dual problem the vector $\nu$ has dimension $d^2$ and is unrestricted. However, for later convenience and w.l.o.g. let us choose the specific form
$
\nu=(\beta \vec{\nu}, \alpha) \, ,
$
where $\alpha \in \mathbb{R}, \beta \in \mathbb{R}^+$ are arbitrary constants and $|\vec{\nu}|=\sqrt{2(d-1)/d}$. From Eqs.~(\ref{dualcond1}) and (\ref{dualcond2}) we have
\begin{eqnarray}
\beta \vec{v}_i \cdot \vec{\nu} + \alpha \leq 0 \, , \\
\alpha > 0 \, .
\end{eqnarray}
A vector $\nu$ will simultaneously satisfy these conditions if and only if $\vec{v}_i \cdot \vec{\nu} < -\alpha/\beta$. We can always choose $\beta$ sufficiently large such that $-\alpha/\beta \rightarrow 0$, so the least restrictive condition has the form
\begin{equation}\label{hemisphere}
\vec{v}_i \cdot \vec{\nu} < 0 
\end{equation}
[taking the complementary equations to (\ref{dualcond1}) and (\ref{dualcond2}) would have led to the equivalent condition $\vec{v}_i \cdot \vec{\nu} > 0$]. To summarize, as long as there exists a vector $\vec{\nu}$ whose scalar product with every other generalized Bloch vector $\vec{v}_i$ is negative, we can always choose two positive constants $\alpha, \beta$ such that $\nu=\left( \beta \vec{\nu},\alpha \right)$ satisfies Eqs.~(\ref{dualcond1}) and (\ref{dualcond2}). Hence, the LP (\ref{LP}) is infeasible and the set of operators $\{E_i\}$ cannot form a POVM. 

Condition (\ref{hemisphere}) has a clear geometrical interpretation: $\vec{\nu}$ defines a hyperplane in $\mathbb{R}^{d^2-1}$ which includes the $\vec{0}$ point and splits a $(d^2-2)$-sphere such that all $\vec{v}_i$ points are situated at one side of the hyperplane. Obviously, if the vectors $\vec{v}_i$ do not span $\mathbb{R}^{d^2-1}$ but a subspace of smaller dimension $d'$, it will suffice to consider hyperplanes of dimension $d'-1$. This hyperplane condition is equivalent to stating that the convex hull of the $\vec{v}_i$ points does not contain the $\vec{0}$ point.

We now state and prove next that, given a POVM with $n>d^2$ non-zero elements, it is always possible to select a subset of at most $d^2$ which is also a POVM, up to a suitable redistribution of weights. This is easily derived from the LP feasibility formulation: equations (\ref{cond3}) and (\ref{cond4}) represent a system of $d^2$ equality conditions and $n$ variables; if such a system is feasible, it would have a single solution for some value of $n\leq d^2$. For $n>d^2$ its solution will have $n-d^2$ extra degrees of freedom, and hence we will always be able to fix $n-d^2$ variables to zero. Since this statement is not valid when $n\leqslant d^2$ (except for the case in which vectors $\vec{v}_i$ span a smaller subspace of $\mathbb{R}^{d^2-1}$), it follows that an extremal POVM will have at most $d^2$ non-zero elements, as it has been noted in previous works \cite{D'Ariano2005,Haapasalo2011}.

The geometrical interpretation of the POVM condition provides a clear and useful picture of the results in the previous paragraph in terms of the distribution of vectors $\vec{v}_i$. Note that the number of vectors needed to subtend a solid angle in $\mathbb{R}^{d^2-1}$ is $d^2-1$. 
The conical hull defined by such vectors contains a portion of a hypersphere $S^{d^2-2}$.
It is then easy to convince oneself that the minimum number of vectors required to cover the whole $S^{d^2-2}$ as a union of conical hulls is $d^2$ [note that such a distribution necessarily implies the violation of condition (\ref{hemisphere}) and, therefore, the fulfilment of (\ref{cond4})].
This means that, given such a set of $d^2$ vectors, if we add an extra vector, it will necessarily fall in a conical hull defined by a certain subset of $d^2-1$ vectors of the original set and thus it could be expressed as a conical combination of those (i.e. as a linear combination with nonnegative coefficients).
Hence, given $d^2+1$ POVM elements whose Bloch vectors satisfy condition (\ref{cond4}), one can always choose one of the vectors and %express it as a convex combination of $d^2-1$ other vectors: 
replace it by a conical combination of $d^2-1$ other vectors:
the remaining set of $d^2$ vectors still satisfies condition (\ref{cond4}).

In general, Bloch vectors $\vec{v}_i$ will be contained in $\mathbb{R}^{d^2-1}$. When $n<d^2$, additional restrictions over vectors $\vec{v}_i$ derive from (\ref{hemisphere}). If $n=2$ then the generalized Bloch vectors $\vec{v}_1$ and $\vec{v}_2$ should span a 1-dimensional space in order to be able to violate condition (\ref{hemisphere}). In fact, the condition is violated only if $\vec{v}_1=-\vec{v}_2$. If $n=3$, vectors $\vec{v}_1, \vec{v}_2$ and $\vec{v}_3$ should lie on a plane and not belong to the same semicircle (defined by a line). For any $n$ we should have 
\begin{equation}
\{\vec{v}_1,\vec{v}_2,\dots,\vec{v}_n\} \in S^{n-2} \subset \mathbb{R}^{n-1} \, ,
\end{equation}
where vectors $\vec{v}_i$ do not belong to any hemisphere of $S^{n-2}$. Note that the extremality statement in the previous paragraph extends to $\mathbb{R}^{n-1}$: if we have $n' \geq n+1$ vectors (whose associated operators form a POVM) that span $\mathbb{R}^{n-1}$, then we can always find subsets of at most $n$ vectors which violate condition (\ref{hemisphere}), and thus are able to form an extremal POVM.

To finish this section and for clarity purposes, note that it has been assumed that the solutions of the LP feasibility problem correspond to extremal POVMs, i.e. extremal points not only of the set of feasible points but also of the set of all POVMs. This is indeed the case: on one hand, such a solution corresponds to a set of linearly independent POVM elements $\{E_i\}$; on the other hand, any POVM with at most $d^2$ rank-1 linearly independent elements is extremal (see e.g. Proposition 3 in \cite{Haapasalo2011}).

\section{The algorithm}\label{algorithm}

In this section, we present our constructive algorithm for decomposing a POVM into extremals. We first address the case of rank-1 POVMs, and then we extend the algorithm to higher-rank cases.
%We now present our algorithm in its general form for rank-1 POVMs, and then we extend it to higher-rank cases. 
We are given a rank-1 POVM $\mathbb{P}_N=\{a_i E_i\}$, $i=1,\ldots,N$, where $E_i$ are normalized operators given by (\ref{elements}) and $a_i>0$. Our aim is to express it as
\begin{equation}\label{decomp}
\mathbb{P}_N=\sum_k p_k \mathbb{P}^{(k)}_n ,
\end{equation}
where $\mathbb{P}^{(k)}_n$ is an extremal rank-1 POVM with $n \leqslant d^2$ outcomes. 
This means that in order to implement $\mathbb{P}_N$ it will suffice to randomly select a value of $k$ from the probability distribution $p_k$, and then perform $\mathbb{P}^{(k)}_n$.
%This means that in order to implement $\mathbb{P}_N$ it will suffice to choose the measurement $\mathbb{P}^{(k)}_n$ with probability $p_k$ and perform it. 
The algorithm we propose to carry out such a decomposition works as follows:

We first define the LP feasibility problem
\begin{eqnarray}\label{LP2}
\mbox{find} &\quad& x \nonumber \\
\mbox{subject to} &\quad& Ax=b \nonumber \\
&\quad& x\geqslant 0 \, ,
\end{eqnarray}
where $x$ is a vector of $N$ variables, $A$ is a matrix whose columns are given by vectors $v_i=(\vec{v}_i,1)$, and $b=(\vec{0},d)$. The set of feasible points of this LP, i.e. the values of $x$ compatible with the conditions of the LP, define a convex polytope $K$ in the space of coefficients:
\begin{equation}
K = \{x \,/\; Ax=b, x\geqslant 0\} \subset \mathbb{R}^N .
\end{equation}
The vertices of $K$ are its extremal points, and the region of $\mathbb{R}^N$ defined by the convex hull of all the vertices contains all the points that can be expressed as convex combinations of these extremal points. Dantzig's {\it simplex method} for solving LPs \cite{Todd2002} starts at a vertex of $K$, and it moves from vertex to vertex minimizing a cost function, until there is no preferred direction of minimization; then, the optimal solution has been found. Since there is no cost function in a feasibility problem, the simplex method applied to (\ref{LP2}) terminates at its first step: when it finds the first vertex. 
The convex polytope $K$ is isomorphic to a subset of $\mathcal{P}_N$, i.e. there is a one-to-one correspondence between all their elements, and they behave equivalently.
Therefore, such a vertex $x^{(1)}=\{x^{(1)}_i\}$ found as the solution of the LP corresponds to the set of coefficients of an extremal POVM, and as such $x^{(1)}$ will have at most $d^2$ and at least $d$ non-zero elements. The vertices of the polytope $K$ correspond to all the extremal POVMs that one can comprise using only the original elements $\{E_i\}$, and its interior region contains all the possible POVMs generated by these extremals.

Once we have found $x^{(1)}$, we algebraically subtract it from the original set of coefficients $\{a_i\}$. To illustrate this operation, let us assume $d=2$ and $x^{(1)}=\{x^{(1)}_1,x^{(1)}_2,0,\ldots,0\}$. Then, $\{a_i\}$ is rewritten as
\begin{eqnarray}\label{1step}
& \{a_1,a_2,a_3,\ldots,a_N\} = p\,x^{(1)} + (1-p) x^{\rm (aux)}, \\
& x^{\rm (aux)}=\left\{\frac{a_1-p\,x^{(1)}_1}{1-p},\frac{a_2-p\,x^{(1)}_2}{1-p},\frac{a_3}{1-p},\ldots,\frac{a_N}{1-p}\right\}.
\end{eqnarray}
For $x^{\rm (aux)}$ to be an element of $K$, the inequality 
\begin{equation}\label{pcond}
p \leqslant a_i/x^{(1)}_i \leqslant 1
\end{equation}
has to hold for all $i$ such that $x^{(1)}_i>0$. To guarantee the left-hand side of (\ref{pcond}), we take
\begin{equation}
p=\min_i \frac{a_i}{x^{(1)}_i}\,.
\end{equation}
Let us reorder the coefficients $\{a_i\}$ and $x^{(1)}$ such that $p=a_1/x^{(1)}_1$. This choice of $p$ makes the first coefficient of $x^{\rm (aux)}$ to be zero (it could happen that more than one element turns to be zero, thus accelerating the algorithm, but we consider from now on the worst case scenario in which one element is eliminated at a time). 
Also, the right-hand side of (\ref{pcond}) is immediately satisfied since $a_1<x^{(1)}_1$.
Note that $p\in\left[0,1\right]$, thus it is a probability.
%Note that $p$ can be regarded as a probability, i.e. $p\in\left[0,1\right]$. 
Now, (\ref{1step}) can be understood as a probabilistic (convex) combination of $x^{(1)}$ and $x^{\rm (aux)}$, both set of coefficients corresponding to an extremal POVM $\mathbb{P}_2^{(1)}$ and a POVM with $N-1$ outcomes $\mathbb{P}^{\rm (aux)}_{N-1}$. Hence, as a result of the first step of the algorithm, we can write
\begin{equation}
\mathbb{P}_N=p\,\mathbb{P}_2^{(1)}+(1-p)\,\mathbb{P}_{N-1}^{\rm (aux)}\,.
\end{equation}
We then repeat this process redefining the LP with $\mathbb{P}_{N-1}^{\rm (aux)}$ as the initial POVM, which gives us another vertex $x^{(2)}$ associated to an extremal POVM with $n$ outcomes $\mathbb{P}^{(2)}_n$, a remainder $\mathbb{P}^{\rm (aux)}_{N-2}$ and its corresponding probabilities. Of course, in general $d\leqslant n\leqslant d^2$. We iterate this process $N-n_L$ times, where $n_L$ is the number of outcomes of the last extremal POVM obtained. At the last step the simplex algorithm will identify a unique solution with probability 1, corresponding to the input set $x^{\rm (aux)}=x^{(N-n_L)}$.

It is important to stress that the polytopes of the LPs at each step of the algorithm, $K^k$, are subsequent subsets of each other, that is 
\begin{equation}
K\supset K^1\supset \ldots \supset K^{N-n_L+1}.
\end{equation}
The result of each step is the elimination of one of the original elements $\{E_i\}$, and with it all the vertices that required that element. 
Thus, each step projects the polytope onto a subspace of the space of coefficients by reducing its dimension by one.
As a consequence, in the end all the vertices selected by the simplex algorithm were vertices of the original $K$.
\\

When the rank of $\mathbb{P}_N$ is higher than 1 we can still apply the same algorithm, just adding two extra steps: one preparation step and one post-processing step. The preparation step works as follows: for every $i$ such that \mbox{$\rank{E_i}>1$}, express $E_i$ in its eigenbasis $\{\ket{v_{ij}}\}$ as
\begin{equation}
E_i=\sum_j \lambda_j \ketbrad{v_{ij}}=\sum_j E_{ij}.
\end{equation}
Consider each rank-1 operator $E_{ij}$ as a new outcome and denote the new (rank-1) POVM by $\mathbb{P}_{\bar{N}}=\{\bar{E}_l\}_{l=1}^{\bar{N}}$, where $\bar{N}=\sum_i \rank{E_i}>N$. The label $l(i,j)$ carries the information contained in labels $i$ and $j$. Now, the algorithm described above can be applied directly over $\mathbb{P}_{\bar{N}}$. The post-processing step is needed for associating the outcomes of the measure finally performed ($l$) to the outcomes of the original $\mathbb{P}_N$ ($i$). 
\\

A generic algorithm for decomposing a point in a convex set into a combination of extremal points of that set can be found in \cite{D'Ariano2005}. Although in this paper D'Ariano {\it et al} specialize it for a general $\mathbb{P} \in \mathcal{P}_N$, we would like to remark that significant differences stand between our algorithm and the one presented there. The algorithm of \cite{D'Ariano2005} consists in a recursive splitting of an inner point of the convex set into a convex combination of two points that lie on a facet of the convex set (and thus a subset of a strictly smaller dimension). After enough steps it yields a number of extremal points along with some weights in a tree-like form, thus statistically reproducing the original point as a mixture of extremal points. The direction in which the splitting is done at each step is determined through an eigenvalue evaluation.
The particular decomposition we have presented in this paper may be considered within this general scheme (we also do binary partitions at each step), however two main differences arise. On one hand, the process of obtaining extremal points (i.e. the direction of splitting) is radically different. We associate a polytope $K$ to a subset of the convex set $\mathcal{P}_N$ via an isomorphism, and then we move efficiently along the directions marked by the vertices of $K$. Thus, there is no need to analyze the whole convex set $\mathcal{P}_N$ (which is strongly convex, i.e. its extremal points are not isolated but lie on a continuum) for a given $\mathbb{P}$: our algorithm does not optimize a direction among a continuum of possibilities at each step but selects any direction of a given finite set. On the other hand, the authors in \cite{D'Ariano2005} state that their algorithm provides a minimal decomposition, with a number of extremals upperbounded by $(N-1) d^2 +1$. We have found that our algorithm yields the tighter bound $(N-1)d+1$.

\section{Ordered decompositions}\label{ordereddecomp}

The algorithm described in Section \ref{algorithm} will produce one of many possible decompositions of the initial POVM into at most $N-n_L+1$ extremals (recall that $n_L$ ranges from $d$ to $d^2$), even if we only consider extremals made of original elements. Because at each step any of the vertices of the polytope could be identified and extracted, the final decomposition obtained is not unique and depends on the particular implementation of the simplex method for solving the LP. That being said, one could be interested in a particular decomposition that exhibits certain properties.
We observe that there is room in our algorithm for these extra requirements while maintaining its structure, that is to efficiently produce decompositions into at most $N-n_L+1$ extremals obtained through a LP solved by the simplex method.
%We observe that, maintaining the structure of our algorithm, there is room for these extra requirements. By maintaining the structure of the algorithm we mean that the decomposition consists of $N-n$ extremals obtained through a LP solved by the simplex method, and thus it is carried out efficiently.
To obtain a particular decomposition with this structure that verifies a certain desired property we will simply have to establish some ranking among the vertices of the polytope in agreement to that property or associated criterion, and tweak the algorithm to choose first the ones at the top of the ranking. This is what we call an {\it ordered} decomposition.

A desirable ordering from the point of view of an experimental realization may be, for instance, to prioritize the vertices with more zero elements, if there is any. Those vertices would correspond to extremals with less outcomes. In the case of $d=2$, for instance, extremal POVMs can have 2, 3 or 4 outcomes. Such a decomposition would seek first for 2-outcome (Stern-Gerlach measurements), then 3-outcome and finally 4-outcome POVMs.

The simplex method is an efficient way of finding the optimal vertex of a polytope according to some criterion, which is 
implemented as
%driven by 
a cost function. This is done by minimizing or maximizing such a cost function. In the description of the algorithm we chose this function to be independent of the variables, because we were only interested in finding a feasible point. The choice of the cost function will vary the direction taken by the simplex algorithm when it moves from one vertex to another, and it is therefore a way to establish a ranking among the vertices. Consider for instance the cost function
\begin{equation}\label{costQ}
Q_n = \sum_{i=1}^n x_i^2 \, .
\end{equation}
The maximization of $Q_n$ on its own could in principle work for finding the vertices with more zeros: if we would have no other constraint but a fixed quantity $d$ to distribute among the $n$ parties $x_i$, the strategy that maximizes $Q_n$ is to give all to one party and zero to the others. But we have more constraints in (\ref{LP2}). Let us take a look on the minimum and maximum values of $Q_4$, that is for extremals with 4 outcomes. The value of $Q_4$ will only depend on the geometric distribution of the outcomes of the extremal. On one hand, $Q_4$ takes its minimum value when $d=\sum_i x_i$ is equally distributed among the variables $x_i$, that is when the 4 associated Bloch vectors $\vec{v}_i$ are orthogonal in pairs (i.e. the POVM is a combination of two Stern-Gerlachs). This value is $Q_4^{\rm min}=(d/4)^2 \times 4=d^2/4$. On the other hand, $Q_4$ reaches its maximum value if three of the vectors are parallel and the fourth is orthogonal to all the others (this is the way to put a maximum weight on one of the $x_i$), that is $Q_4^{\rm max}=(d/2)^2+(d/6)^2\times 3=d^2/3$. Applying the same reasoning for 3-outcome extremals we have $Q_3^{\rm min}=d^2/3$ and $Q_3^{\rm max}=3d^2/8$, and 2-outcomes can only give $Q_2=d^2/2$. Since
\begin{equation}\label{Qd2}
Q_2>Q_3^{\rm max} > Q_3^{\rm min} = Q_4^{\rm max} > Q_4^{\rm min} \, ,
\end{equation}
the maximization of function $Q_n$ prioritizes the extremals with fewer outcomes at least for $d=2$, when the maximum number of nonzero elements in a vertex is $n=4$.
This, unfortunately, stops being valid for $n>4$, which in general happens if $d>2$. 

The general problem of maximizing a convex function over a convex set of feasible points is called {\it convex maximization}. The problem at hand belongs to this category. While the more standard class of {\it convex minimization} problems (i.e. minimizing a convex function over a convex polytope) count on efficient solving algorithms, this is not the case for convex maximization, except for very special cases. The efficiency of the convex minimization relies on the uniqueness of the convex function's minimum, which is an inner point of the polytope. Conversely, its maxima are located on the vertices of the polytope and all but one are \textit{local} maxima. This fact makes the convex maximization problems intractable in general, and so it is the maximization of (\ref{costQ}). The difficulty lies on the fact that an algorithm might find a local maximum (a vertex), but there is no way to certificate its global optimality (although there are algorithms that, despite no proof certificate, provide good guesses \cite{Fortin2010}). 

Any global search algorithm (able to guarantee global optimality) for convex maximization somehow {\it enumerates} all the vertices, and thus its efficiency highly depends on the number of those. Of course, the ordered decomposition we are looking for is immediately obtained if one enumerates all the vertices of $K$. With such a list, we would just have to pick up first those vertices with more zero elements, corresponding to the extremals with fewer outcomes (or according to any other criterion we may wish). Furthermore, no additional optimization is required since we can extract from the same list the vertex required at each step, thus keeping us from solving a LP for doing so.
The problem of enumerating the vertices of a bounded polyhedron is NP hard in the general case \cite{Khachiyan2008}, but has efficient algorithms able to generate all vertices in polynomial time (typically linear in the number of vertices) for several special cases. For instance, in \cite{Avis1992} there is an algorithm that enumerates the $v$ vertices of a convex polyhedron in $\mathbb{R}^m$ defined by a system of $D$ linear inequalities in time $O(mDv)$. Our polytope $K$ is of this type, and hence we could use the algorithm for our purpose. Note however that $v$ has a direct dependence on $m$ and $D$. The problem of computing $v$ for a given polytope is NP-hard, but a bound can be provided \cite{Barvinok2011}: the number of vertices of our polytope $K\subset\mathbb{R}^m$ is at least exponential in $m$.

In summary, an ordered decomposition of a POVM can be carried out in two ways. On one hand, nonlinear programming techniques can be used to maximize a cost function subject to the constraints of (\ref{LP2}), but none of them will perform with perfect accuracy. We have found a cost function that prioritizes the extremals with less outcomes for $d=2$, but not for greater dimensions. Finding a cost function is problem-specific, and it seems to be highly non-trivial: its maximization should lead first to a vertex of the polytope, and secondly it should move from one to another maximizing the desired property. On the other hand, an alternative method is to enumerate all the vertices of the polytope $K$ defined by the constraints of (\ref{LP2}), but the number of vertices and thus the time required to carry out the enumeration grows exponentially with the number of elements of the original POVM.

\section{Conclusions}

We have presented an efficient algorithm to decompose any POVM $\mathbb{P} \in \mathcal{P}_N$ into extremal ones. The decomposition achieved consists of a convex combination of at least $N-n_L+1$ (if $\mathbb{P}$ is rank-1) and at most $Nd-n_L+1$ (if $\mathbb{P}$ is full-rank) extremal measurements, where $n_L$ ranges from $d$ to $d^2$ and its value is determined by each particular $\mathbb{P}$. In the case in which $\mathbb{P}$ presents some symmetry (as the BB84 POVM shown as an example in Section \ref{notation}), more than one element may be eliminated in one step of the algorithm and thus the number of extremals would be even less. We have shown that only extremal rank-1 POVMs are required to effectively implement $\mathbb{P}$ by introducing a classical post-processing of the outcomes. The decomposition is efficiently carried out by an algorithm based on resolutions of LPs using the simplex method, within polynomial time in $N$ and $d$. The efficiency is achieved by restricting the analysis to a polytope-shaped subset of $\mathcal{P}_N$ for a given $\mathbb{P}$, and thus by taking into consideration only a finite number of extremals (the vertices of the polytope), in contrast to what other authors have considered so far (see e.g.  \cite{D'Ariano2005}). Furthermore, in \cite{D'Ariano2005}, a generic decomposition algorithm that yields a certain maximum number of extremals is provided. We have found that our algorithm beats this performance in a worst case scenario.

Since a given POVM admits many decompositions, we also explore the possibility of obtaining a particular decomposition that exhibits a certain desired property, introduced in the algorithm as an input. We call these decompositions \textit{ordered}, and they are based on prioritizations of extremals that can be made out of subsets of the elements of $\mathbb{P}$. As an example we give a method to prioritize extremal POVMs with less outcomes in the case of $d=2$, and show that either efficiency or accuracy necessarily get compromised.

\section*{ACKNOWLEDGEMENTS}

GS and BG acknowledge financial support from ERDF: European Regional Development Fund. This research was supported by the Spanish MINECO, through contract FIS2008-01236 and FPI Grant No. BES-2009-028117 (GS); and the Generalitat de
Catalunya CIRIT, contract  2009SGR-0985. ACD and SDB were supported by the ARC via the Centre of Excellence in Engineered Quantum Systems (EQuS), project number CE110001013.
SDB thanks Robin Blume-Kohout and Rob Spekkens for extended discussions on the convex structure of POVMs. The methods proposed in this paper are inspired by a constructive but less efficient method to decompose POVMs into extremal elements developed by Robin Blume-Kohout and SDB (unpublished), independently of the existence proofs for such decompositions given in Refs. \cite{D'Ariano2005,Chiribella2007b}. GS and BG thank John Calsamiglia, Emili Bagan and specially Ramon Mu\~{n}oz-Tapia for useful discussions.

\bibliographystyle{plain}

\end{document}